\documentclass[aps,amsmath,amssymb,prl,twocolumn,reprint,showpacs]{revtex4-1}

\usepackage{graphicx}
\usepackage{dcolumn}
\usepackage{bm}
\usepackage{subfigure}
\usepackage{float}
\usepackage{color}
\usepackage{soul}

\begin{document}

\title{Hyperparametric effects in a whispering-gallery mode rutile dielectric resonator at liquid helium temperatures}
\author{Nitin R.~Nand*, Maxim Goryachev, Jean-Michel Le Floch, Daniel L. Creedon, and  Michael E. Tobar}
\affiliation{ARC Centre for Engineered Quantum Systems, School of Physics, The University of Western Australia, 35 Stirling Highway, Crawley 6009, Western Australia,\\*now at the ARC Centre for Engineered Quantum Systems, Department of Physics and Astronomy, Macquarie University, 1 Balaclava Road, North Ryde 2109, NSW, Australia}

\date{\today}

\begin{abstract}
We report the first observation of low power drive level sensitivity, hyperparametric amplification, and single-mode hyperparametric oscillations in a dielectric rutile whispering-gallery mode resonator at 4.2 K. The latter gives rise to a comb of sidebands at $19.756$ GHz. Whereas most frequency combs in the literature have been observed in optical systems using an ensemble of equally spaced modes in microresonators or fibers, the present work represents generation of a frequency comb using only a single-mode. The experimental observations are explained by an additional 1/2 degree-of-freedom originating from an intrinsic material nonlinearity at optical frequencies, which affects the microwave properties due to the extremely low loss of rutile. Using a model based on lumped circuits, we demonstrate that the resonance between the photonic and material 1/2 degree-of-freedom, is responsible for the hyperparametric energy transfer in the system. 
\end{abstract}

\maketitle

\section{Introduction}

Rutile (TiO$_{2}$) is a centrosymmetric tetragonal crystal which is a useful and widely studied material in science and engineering. The crystal is a wide band gap semiconductor, and demonstrates favorable optical properties, such as high refractive index, birefringence, and transparency to wavelengths $\geq$ 400 nm. These properties make rutile suitable for nonlinear applications, such as ultra fast switching, logic, and wavelength conversion~\cite{evans2012}. Rutile is well known to exhibit a host of nonlinear phenomena including polaritonic~\cite{polyarit, polyarit2} and polaronic effects~\cite{polar}, and near `ferroelectric catastrophe' behavior~\cite{ferro}. Third harmonic generation in particular, has been investigated in numerous publications~\cite{nonlinear1, Greenberger1990, fabien2006, fabien2007, benceikh2007, borne2012}. Many of these works~\cite{polyarit, polyarit2, nonlinear1} characterize an effective third order susceptibility $\chi^{(3)}$ of the rutile crystal which leads to four wave mixing, but unlike the aforementioned effects, is always observed at optical frequencies.

At microwave frequencies, the linear dielectric properties of rutile have been well studied, however rather little work has been published on nonlinear effects in this frequency regime. In much of the previous work the rutile crystal has been used as a whispering-gallery (WG) mode resonator~\cite{tobarrutile1998, tobarrutile2001,luitenrutile1998} below $5$~GHz. Rutile is a particularly interesting low loss material due to its high uniaxial anisotropic permittivity, an order of magnitude higher than that of sapphire, and its negative temperature coefficient of permittivity. Thus, WG mode resonators of much smaller size and higher frequency can be made from rutile compared to sapphire. By virtue of its low microwave losses, very high quality factor ($Q$) WG modes can be supported in such resonators at cryogenic temperatures, where electromagnetic energy is strongly confined around the inner surface of the resonator at the interface of the crystal medium and external environment due to total internal reflection.

Over the past few years, WG mode resonators have featured in numerous publications describing the generation of frequency combs~\cite{nature2012}. Many applications including optical spectroscopy, precision frequency metrology, the development of optical clocks, and precise calibration and measurement in observational astronomy all rely on such combs. Optical hyperparametric oscillations in CaF$_{2}$ resonators have been reported~\cite{savchenkov2004} and theoretically analyzed~\cite{matsko2005, chembo2010}, and comb generation has been seen in a range of crystals, such as CaF$_2$~\cite{grudinin2009, savchenkov2008}, and silica toroidal microcavities~\cite{delhaye2007}. More recently, frequency comb generation via four wave mixing in MgF$_{2}$ resonators was reported~\cite{herr2011,grudinin2012}, and comb generation has begun to extend into the mid-infrared regime~\cite{nature2012}.

In the present work, we describe new experimental results and show that a nonlinear phenomenological model possessing $1\frac{1}{2}$ degrees-of-freedom explain the results. The single degree-of-freedom represents the WG mode resonance at the frequency of interest, while the additional half degree-of-freedom represents a resonance highly detuned from the optical domain, which defines the losses at microwave frequencies and possesses nonlinear losses. The model predicts three new types of nonlinear behavior that generally do not exist for nonlinear systems, such as Duffing or Van der Pol oscillators. Single-mode hyperparametric oscillations are observed, which gives rise to a comb of sidebands and low power overdamping. All of these effects are simultaneously observed in the WG mode resonator as predicted by the model, and hence we rule out the causes due to temperature, paramagnetic ions, semiconductor, and even electron spin resonance effects.


Most of the existing literature on nonlinear resonant systems, such as WG mode resonators employs simple one-dimensional models, usually related in some way to the well studied Duffing or Van der Pol oscillators. The former model is a damped oscillator with a nonlinear angular frequency, whereas the latter is a single degree-of-freedom oscillator with a nonlinear loss term. In the case of optical WG resonators, nonlinear effects are modeled as nonlinear correction terms for the refractive index or susceptibility, proportional to the square of the intensity of the electric field~\cite{kerr, Chembo}, or by direct inclusion of nonlinear interaction terms into a Hamiltonian of interacting modes and the corresponding Langevin equations~\cite{matsko2005}. Nonlinear behavior in such systems is well studied and is observed usually in the form of a \textit{pitchfork} bifurcation, excess losses at high powers, or hyperparametric oscillations in many-mode systems. Nevertheless, the number of nonlinear phenomena in resonant systems can be enriched simply by increasing the system degree-of-freedom by one half. In this section, we analytically show that such a system can lead to three new phenomena; low power overdamping; single-mode hyperparametric oscillations; and hyperparametric amplification.

\section{Model}

\begin{figure}[!t]
\centering
\includegraphics[width=0.43\textwidth]{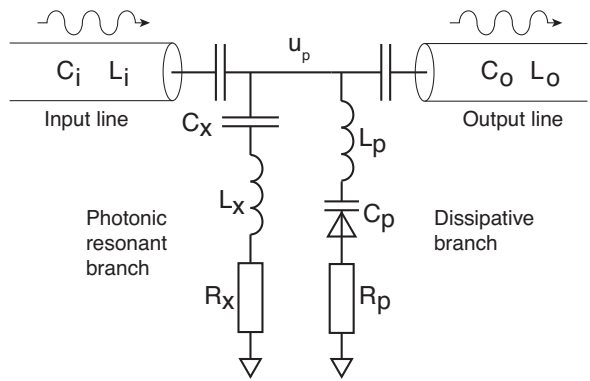}
\caption{An equivalent lumped circuit model representing a nonlinear $1\frac{1}{2}$ degree-of-freedom system, with a photonic WG mode ($C_x$, $R_x$, $L_x$) and the material dissipative half degree-of-freedom ($C_p$, $R_p, L_p$).}
\label{model}
\end{figure}

The model consists of two branches; a resonant photonic branch representing a WG mode, and a dissipative nonlinear branch representing an intrinsic material degree-of-freedom, the optical phonon mode. The latter is taken to be highly detuned, so that its overall response could be represented only by a half degree-of-freedom (or a single reactance). Thus, in terms of an equivalent circuit (see Fig.~\ref{model}), the system is modeled by a resonant series RLC circuit (the photonic branch) in parallel to the material branch consisting of a nonlinear capacitance with dissipation. The effective inductance can be ignored due to the high detuning of the material resonance from optical frequencies. In mathematical language, this representation is a degenerate case of a class of models represented by two coupled nonlinear oscillators. For simplicity, we include in the description just the nonlinearity due to the material behavior, which means only the half degree-of-freedom is assumed to be nonlinear.

A Hamiltonian in terms of the variables of the equivalent circuit model may be written as
\begin{equation}
\label{A033DF}
\left. \begin{array}{ll}
\displaystyle H =  \frac{1}{2L_x}\phi_x^2+\frac{1}{2C_x}q_x^2 + \frac{1}{2L_p}\phi_p^2+\frac{1}{2C_p}q_p^2+ \frac{L_p\eta}{4}\eta q_p^4,\\
\displaystyle \approx \frac{1}{2L_x}\phi_x^2+\frac{1}{2C_x}q_x^2+ \frac{\eta L_p}{4} q_p^4,
\end{array} \right. 
\end{equation}
where $q_x$ is a charge in the resonant photonic branch, $q_p$ is a charge in the nonlinear loss branch, $V_{inp}$ is the driving signal (the incident traveling wave in the transmission line at the point of the resonator), $\gamma_x$ and $\omega_x$ are the damping coefficient and angular frequency of the photonic branch, $\gamma_c$ is a coupling coefficient, and $\eta$ is the coefficient of the nonlinear term derived from the voltage-charge relation of the nonlinear capacitance $u_p = \eta q_p^{3}-\phi$. The approximation only holds for strong nonlinearity. The $C_x-L_x$ pair represents the photonic mode resonance, and $C_p-L_p$ refers to the optical phonon mode. Additional dissipative elements in Fig.~\ref{model} are due to the dissipative bath $\sum \omega_i$. 
The equations of motion for the equivalent system are written in terms of the charge in two capacitances, such that
\begin{equation}
\left\{ \begin{array}{rl}
\displaystyle \ddot{q}_x + 2\gamma_x \dot{q}_x + \omega_x^2 q_x + 2\gamma_c\dot{q}_p &= V_{inp}\\
\displaystyle 2\gamma_p \dot{q}_p + \eta q_p^3 + 2\gamma_c\dot{q}_x &= V_{inp}
\end{array} \right. 
 \label{H004HS}
\end{equation}
Note that the Hamiltonian corresponding to the equations of motion (Eq. \ref{H004HS}) could be written using only three operators; flux $\phi_x$ and charge $q_x$ for the photonic mode, and only charge for the material mode $q_p$.

It can be shown analytically that in the linear limit ($\eta q_p^3\rightarrow 0$), the total frequency response of the system ($q_x(\omega)+q_p(\omega)$) can be made totally independent of detuning frequency ($\omega_x-\omega$) in the limit $\gamma_p\rightarrow\gamma_c$. This is equivalent to the effective nonexistence of the mode at low powers (see Figs.~\ref{intensityplot} and \ref{intensityplot-b}). Although when nonlinearity exists in the second branch, this leads to a situation when external pump photons could be `damped' by the dissipative branch, or stored and re-emitted from the resonant photon branch depending on their number. Only when the dissipative branch is saturated can the photons be transmitted through the mode resonance. 

\begin{figure}[!h]
	\begin{center}
	\includegraphics[width=0.50\textwidth]{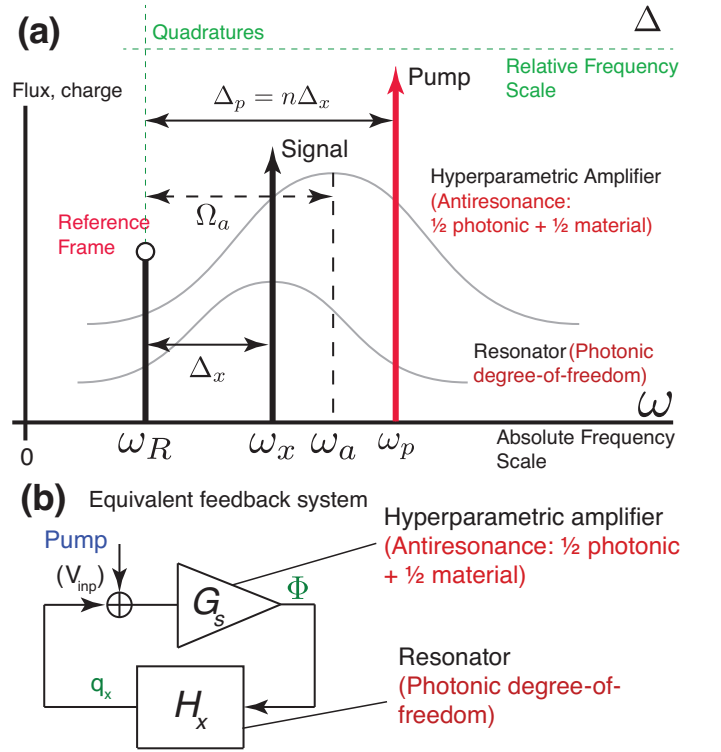}
	\caption{(a) Relationship between the frequencies of the hyperparametric system: the introduction of a reference frequency transforms the problem into a pure parametric one in the relative frequency scale, where pump frequency is an integer multiple of signal. (b) Equivalent block diagram of the system representing resonant and antiresonant behavior in a closed-loop feedback system.}
		\label{figA5}
	\end{center}
\end{figure}

We attribute the notion of a hyperparametric process to the process of energy transfer in a nonlinear system between frequencies whose difference is much less than the frequencies themselves. This notion was first introduced in the domain of optical comb generation~\cite{grudinin2009, savchenkov2008, nature2012}.
In the case of parametric oscillators, the energy of the pump will sustain oscillations when it is an integer multiple of the oscillation frequency, i.e. pump frequency of $n\omega$, where $n$ is an integer and $\omega$ the oscillation frequency. Under hyperparametric oscillation, energy is transferred from the pump $\omega$ into its sidebands $\omega\pm n\Delta\omega$, where $n$ is an integer. 
Although some of the literature on optical comb generation refers to these effects as parametric, we use the term `hyperparametric' to emphasize the unique situation in which the energy transfer is made between signals with a frequency spacing on the order of one mode bandwidth. The hyperparametric problem can be reduced to the usual parametric case simply by transforming the problem from the domain of actual signals into a domain of its quadratures by applying the method of slowly varying amplitudes relative to some reference frequency (see Fig.~\ref{figA5}(a)).

The $1\frac{1}{2}$ degrees-of-freedom system exhibits two types of resonant behavior, the first being a series resonance (a pure resonance of the photonic degree-of-freedom) between the inductive element $L_x$ and capacitive element $C_x$. The second resonant behavior is antiresonance (mixed resonance of the photonic half-degree-of-freedom and the material half-degree-of-freedom), which is a parallel resonance between $C_x$ and the inductance $L_p$ in the nonlinear branch. The notion of antiresonance is commonly used in the acoustic device community where a series mechanical resonance is always accompanied by an antiresonance - parallel resonance with capacitive electrodes. In the present work, it is the antiresonance behavior that is responsible for hyperparametric energy transfer. To demonstrate this, a corresponding equation of motion can be written in terms of flux, such that 
\begin{equation}
\displaystyle \frac{\phi^{\frac{2}{3}}}{\Big(\dot\Phi-\phi\Big)^{\frac{2}{3}}}{\ddot\Phi}+2\delta_a\dot\Phi+\omega_a^2\Phi = k_v V_{inp}+k_p\dot{q}_x,\\
 \label{F014GG}
\end{equation}
where $2\delta_a = {3}Z_i{\eta^{\frac{1}{3}}}{\phi^{-\frac{2}{3}}}$, $\omega_a^2 = \frac{3}{L_p}\eta^\frac{1}{3}{\phi^{-\frac{2}{3}}}$, $k_v= 3\eta^\frac{1}{3}{\phi^{-\frac{2}{3}}}$, $k_p = 3\eta^\frac{1}{3}\frac{R_x}{L_p}{\phi^{-\frac{2}{3}}}$, $\dot{\Phi} = u_p$, and $q_x$ is a charge from the pure photonic mode. The flux calculated using this equation of motion is fed back into the resonant part creating a closed-loop feedback system (see Fig.~\ref{figA5}(b)).

The nonlinear subsystem described by Eq. \ref{F014GG} exhibits two types of external forcing: $V_{inp}$ due to the external pump signal, and $\dot{q}_x$ which is a signal coming out of the pure photonic subsystem. To demonstrate the energy transfer from the former to the latter, we apply the method of slowly varying amplitudes~\cite{krylov01, krylov02} (a rotating wave approximation) assuming forcing terms of the form
\begin{equation}
\left. \begin{array}{rl}
\displaystyle V_{inp} &= V_0\sin\omega_pt,\\
\displaystyle q_x &= A\sin\omega_xt+B\cos\omega_xt=q_{x0}\cos\big(\omega_xt + \varphi\big),\\
\end{array} \right. 
 \label{F019GG}
\end{equation}
where $\omega_p$ and $\omega_x$ are angular frequencies of the pump and signal in the absolute frequency scale.
The solution is then represented in the form
\begin{equation}
\left. \begin{array}{ll}
\displaystyle \Phi=U\cos\omega_Rt+V\sin\omega_Rt,\\
\end{array} \right. 
 \label{F020GG}
\end{equation}
where $\omega_R$ is some reference frequency. It is chosen in such a way that 
\begin{equation}
n\Delta_x = n(\omega_x-\omega_R) = \omega_p-\omega_R = \Delta_p,
 \label{F021GG}
\end{equation}
where $n\in\mathbb{Z}$, $\Delta_p$, and $\Delta_x$ are angular frequencies of the pump and signal relative to the reference frequency $\omega_R$.
Thus, in the new relative frequency scale, the pump frequency is an integer ($n$) multiple of the signal frequency. The relationships between reference frequencies and absolute frequencies are shown in Fig.~\ref{figA5}. Thus, Eq. \ref{F020GG} transforms the initial hyperparametric problem in terms of flux $\Phi$ and charge $q_x$ into an $n^{\text{th}}$ order parametric problem in terms of quadratures $U$ and $V$.

The $n^{\text{th}}$ order parametric problem corresponding to the hyperparametric amplifier under analysis can be written as
\begin{equation}
\left\{ \begin{array}{ll}
\displaystyle \dot{V} = -\Big[\Omega_a-\mu\big(3V^2-U^2\big)\Big]U-\delta_a\Big[1-\chi\big(V^2+U^2\big)\Big] V+\\
\displaystyle\hspace{25pt} +\frac{k_v}{2\omega_R}V_0\sin\Delta_p+
 \frac{k_p\omega_x}{2\omega_R} q_{x0}\sin\big(\Delta_x+\varphi\big),\\
\displaystyle \dot{U} = -\delta_a\Big[1 -\chi\big(V^2+U^2\big)\Big]U+\Big[\Omega_a-\mu\big(3U^2-V^2\big)\Big]V-\\
\displaystyle\hspace{25pt}-\frac{k_v}{2\omega_R}V_0\cos\Delta_p - \frac{k_p\omega_x}{2\omega_R}q_{x0} \cos\big(\Delta_x+\varphi\big),\\
\end{array} \right. 
 \label{F022GG}
\end{equation}
where $\Omega_a = \frac{\omega_a^2-\omega_R^2}{2\omega_R}$, $\chi = \frac{15}{9\cdot 8}\omega^2\phi^{-2}$, and $\mu=\chi\frac{\omega_a^2}{\omega}$. It is assumed that $\dot{V}\ll U\omega_R$ and $\dot{U}\ll V\omega_R$, i.e. the reference frequency is much larger than $\Delta_x$ and $\Delta_p$. Also the system is considered to be low loss, i.e. $\omega_R\gg\delta_a$. This approximation allows us to neglect terms appearing in Eq. \ref{F022GG} associated with multiplication of $k_vV_{inp}$ and $k_p\dot{q}_{x}$ terms by ${\Big(\dot\Phi-\phi\Big)^{\frac{2}{3}}}$ terms in Eq. \ref{F014GG}. Hence, we consider the case
\begin{equation}
\delta_a\sim\chi\Upsilon^2, \hspace{10pt} \omega_R\gg\chi\Upsilon^2
 \label{F022GGa}
\end{equation}
where $\Upsilon$ is both $U$ or $V$. This is a very important advantage of the hyperparametric amplifier in comparison with a parametric system. In the former case, the nonlinearity resulting in a parametric process must be of the order of mode losses ($\delta_a\sim\chi\Upsilon^2$), whereas in the latter case it must be comparable with the mode frequency itself $\omega_R\sim\chi\Upsilon^2$. 

\begin{figure}[!h]
	\begin{center}
	\includegraphics[width=0.48\textwidth]{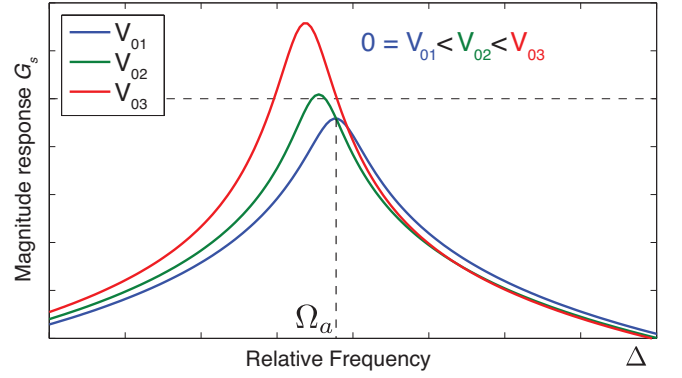}
	\caption{Small signal magnitude response $G_s(\Delta)= \frac{\sqrt{\widetilde{u}^2+\widetilde{v}^2}}{{q}_{x0}}$ of the hyperparametric amplifier in the quadrature space for different values of the pump power, where $\widetilde{u}$ and $\widetilde{v}$ are quadratures of the system response to small signal $q_{x0}$. }
		\label{figA67}
	\end{center}
\end{figure}

The small signal gain of the system described by Eq. \ref{F022GG} for the internal resonator signal $q_x$ is calculated numerically using the harmonic balance method, up to the 5th harmonic. The result is shown in Fig.~\ref{figA67}. The figure demonstrates amplification of the signal from the resonator part of the system. Since both parts are interconnected into a feedback system, the entire closed-loop system becomes unstable when the loop amplification exceeds unity and phase matching conditions are met. In this situation, the parametric amplifier part of the system can compensate the losses in the resonant part. Such regimes are predicted by the numerical simulations. This situation is a regime of sustainable parametric oscillations in the quadrature space of relative frequency that corresponds to hyperparametric sideband generation in the original space. 

In the degenerate regime ($n=0$), the system described by Eq. \ref{F022GG} is able to transfer energy from the strong pump to a signal degenerate with it in frequency, but associated with different input source, e.g. noise generated by $R_x$. In this case, the system behaves as a hyperparametric degenerate amplifier, so that the noise signal of a frequency coming into the resonator would be amplified, and then read out upon transmission.  

In summary, the method of slowly varying amplitudes was applied to the original problem producing hyperparametric oscillations in the domain of actual real physical quantities, with the signal represented by a set of coupled nonlinear differential equations. The order of the problem is one and a half due to large detuning of the material degree-of-freedom. The applied technique transforms the original system from this domain into a domain of oscillating quadratures in the relative frequency frame. The system of nonlinear differential equations which we derived can be considered itself as a nonlinear problem relating quantities in the quadrature domain. The systems output signal in the original domain is a comb, a set of equally spaced ($\Delta_x$) spectral lines near the reference frequency ($\omega_R$). The energy is transferred through a hyperparametric process from the pump into the sidebands. In the quadrature domain, there are several signals of the frequencies $k\Delta_x$, where $k$ is an integer. The energy is transferred from the pump that corresponds to one of them through the usual parametric process. Thus, hyperparametric problems in the domain of actual physical quantities may be approximately reduced to a parametric problem in the domain of quadratures. 

\section{Experimental observations}

Whispering-gallery modes in the frequency range $9$ to $22$~GHz with $Q$-factors of a few million were excited in a cylindrical rutile crystal of diameter $14.08$~mm and height $5.57$~mm. The crystal was supported inside a copper cavity (internal diameter $21.06$~mm and height $9.01$~mm) using two teflon supports via a thin capillary bored through the center of the rutile specimen. The cavity was cooled inside a cryocooler with the temperature regulated near $4.2$K. Two loop probes were used to couple to the resonator, one to excite WGE modes (coupling to the $E_{r}$ field component) and the other, a detection probe placed at 180$^{\circ}$ to the excitation probe. The resonator was characterized in transmission using a vector network analyzer (VNA), allowing mode characterization over a wide range of frequencies and incident powers. The coupling coefficient of the probes was set low so that the measured modes were undercoupled. In this way, the measured properties (such as $Q$) are close to their intrinsic values, and effects such as loading of the Q-factor due to the measurement apparatus are minimized. 

By varying the input signal power, three distinct types of nonlinear behavior have been observed in the rutile crystal; a) low power overdamping; b) single-mode hyperparametric oscillations; and b) hyperparametric 
amplification which are explained in the following paragraphs. These nonlinear features are common for every mode. 

Firstly, a drive level sensitivity or excess loss is observed at low driving power. Below a certain lower power threshold, or in what is supposed to be a linear limit, mode profiles are unmeasurable as if the mode has vanished. This effect is clearly seen in Figs.~\ref{intensityplot} and \ref{intensityplot-b}, where modes do not exist below $-12$ and $-25$~dBm correspondingly. The model presented in the previous section explains this behavior, which cannot be demonstrated for a model with one degree-of-freedom without introducing a singularity.

Secondly, the mode vanishes at high drive power levels which defines an upper threshold. This effect is also observed in Fig.~\ref{intensityplot} at approximately $7$~dBm. Such observations can be made for a Duffing-like model when the nonlinearity dominates over linear frequency selective behavior. In other words, when the nonlinear term in the equation of motion becomes larger than the linear dynamic term.
 
\begin{figure}[!t]
\centering
\includegraphics[width=0.47\textwidth]{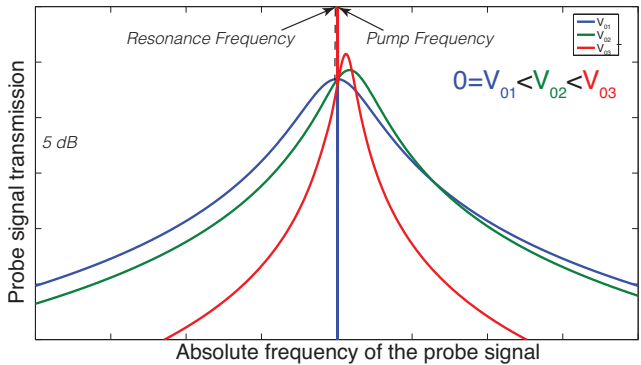}
\caption{Numerical calculation of the transmission coefficient of the small probe signal through the system in the presence of the slightly detuned pump.}
\label{erre}
\end{figure}

\begin{figure}[!t]
\centering
\includegraphics[width=0.47\textwidth]{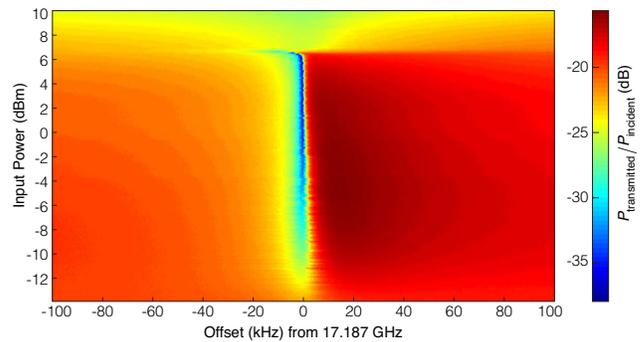}
\caption{Intensity plot of the ratio of transmitted signal to incident signal for a typical WG mode around the center frequency of $17.2$~GHz for different power levels of the input signal. The mode demonstrates low drive level sensitivity (disappears below $-12$~dBm), and strong nonlinear effects at high powers.}
\label{intensityplot}
\end{figure}

\begin{figure}[!h]
\centering
\includegraphics[width=0.48\textwidth]{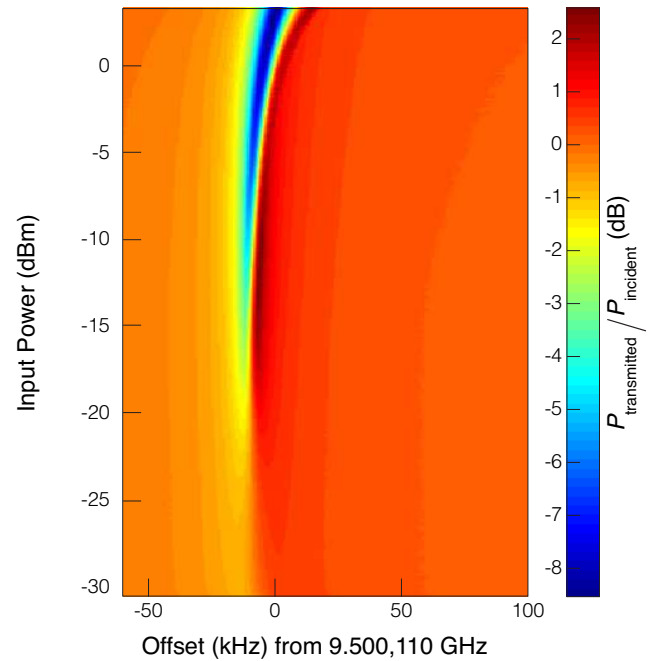}
\caption{Intensity plot of the ratio of a transmitted signal to incident signal for a typical WG mode around the central frequency of $9.5$~GHz for different power levels of the input signal. The mode demonstrates low drive level sensitivity.}
\label{intensityplot-b}
\end{figure}

Thirdly, we demonstrated that when a weak probe signal is tuned on resonance, it exhibits amplification in the presence of a strong detuned pump. The effect exists only within the mode bandwidth and significantly weakens when the probe is being detuned from the resonance. This amplification has been simulated numerically using the model in its original charge representation by the Harmonic balance method (see Fig.~\ref{erre}).

\begin{figure}[b]
\centering
\includegraphics[width=0.48\textwidth]{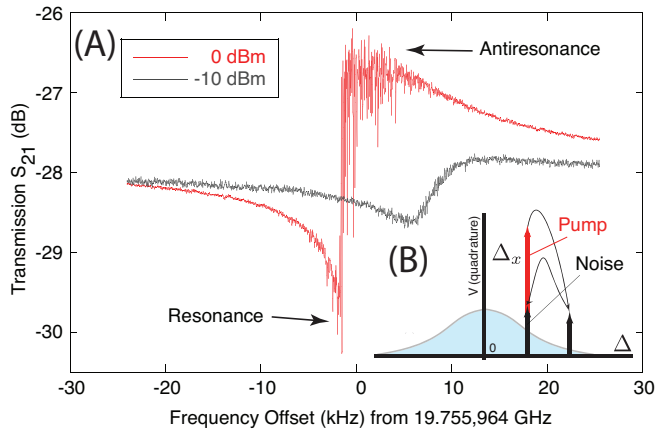}
\caption{(a) The WG $19.756$ GHz mode showing a power effect for input powers of 0 dBm and -10 dBm. (b) The energy transfer diagram for the degenerate ($n=0$) internal noise amplification process.}
\label{fig3}
\end{figure}

Finally, as a manifestation of the previous effect, the excess noise near the discontinuity is explained by hyperparametric amplification of the noise (see Fig.~\ref{fig3}). This effect is also explained by the $1\frac{1}{2}$ degrees-of-freedom model. It corresponds to a degenerate regime ($n=0$) for small $\Delta_x$ when the detected signal contains an amount of system noise hyperparametrically amplified by an incident pump. This effect has been numerically confirmed by calculating a small signal transfer function from the resonator internal noise source to the output.

\section{Microwave hyperparametric oscillations}

Another manifestation of nonlinear phenomena in the rutile WG mode resonator is hyperparametric single-mode microwave oscillation.
This type of oscillation was observed as sideband frequencies generated around the transmitted signal when the input pump frequency was tuned to near that of a WG mode at $19.756$~GHz. The experimental setup consisted of a microwave synthesizer, a cryocooled rutile resonator, and a spectrum analyzer. 

\begin{figure}[t]
\centering
\includegraphics[width=0.48\textwidth]{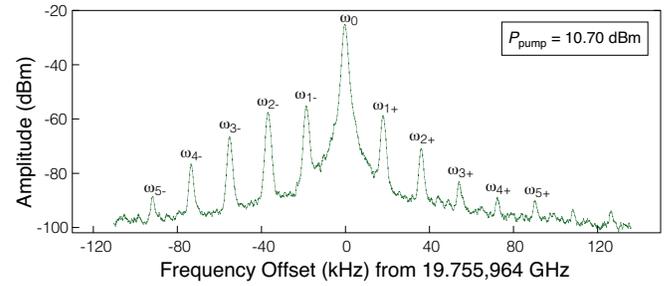}
\caption{Spectrum analyzer trace of the sidebands generated in rutile with a WGE mode at $19.756$ GHz. The comb spacing is $20$~kHz at the given power and proportional to the pump frequency detuning.}
\label{fig6}
\end{figure}

A typical spectrum of the transmitted signal is shown in Fig.~\ref{fig6}. The oscillations could only be excited for a certain range of input powers and detuning frequencies (several tens of kilohertz blue-detuned from the WG resonance frequency). None of the additional tones coincided with any other WGM. The closest WGM is several orders of magnitude away compared to the frequency range analyzed. The system demonstrated an upper and lower threshold power (16 mW and 8 mW respectively) at which the sidebands totally disappeared.

\begin{figure}[t]
\centering
\includegraphics[width=0.48\textwidth]{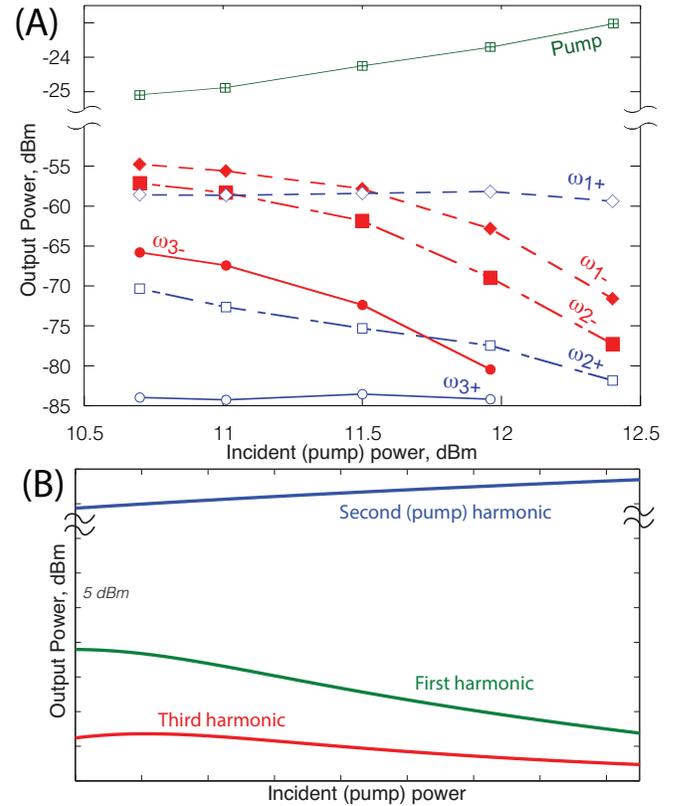}
\caption{(a) Experimental observation of power dependance of sidebands as labeled in Fig. \ref{fig6}. (b) Numerical calculation with the proposed model showing the decrease of generated signal power with increasing pump power.} 
\label{fig8}
\end{figure}

The intensity of the sidebands strongly depended on the power and detuning of the input signal (see Fig.~\ref{fig8}). The intensity decreased with increasing pump power. This effect has been predicted by numerical simulation.
As is predicted by the theory, the comb spacing $\Delta_x$ is the pump detuning from the resonance, such that
\begin{equation}
\Delta_x = \frac{\omega_p-\omega_x}{n-1}.
 \label{F022DFa}
\end{equation} 
The power dependence of the comb spacing $\Delta_x$ is due to the power dependence of the $\omega_x$ component in Eq. \ref{F022DFa} that appears due to Kerr nonlinearity of the photonic mode, and nonlinear effects of higher order. Variation in the comb spacing over such a wide range implies that the effect does not depend on some characteristically well-defined frequency, such as those expected for mechanical vibrations. 

\begin{figure*}[t]
\centering
\includegraphics[width=\textwidth]{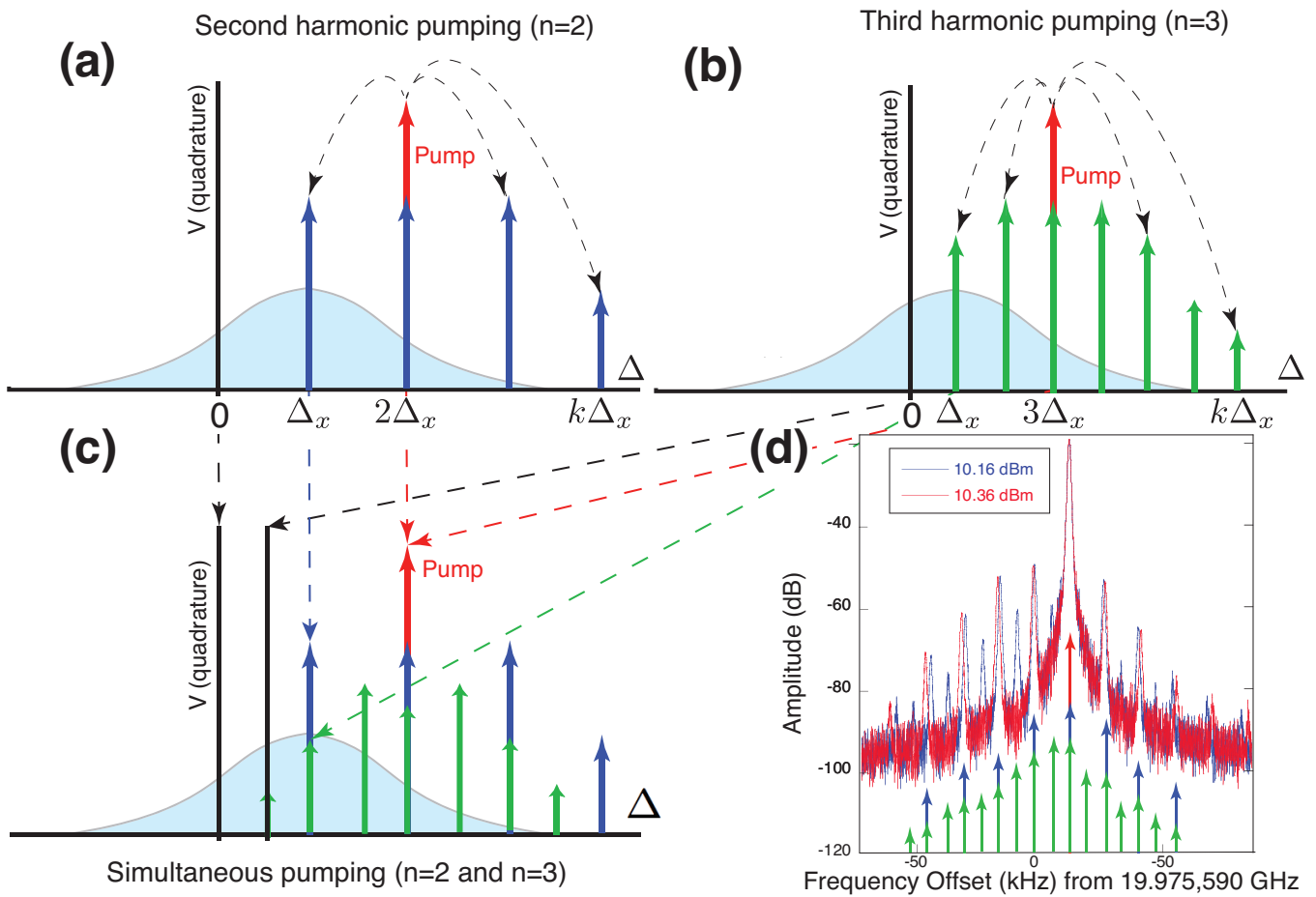}
\caption{Effect of co-existence of different solutions: (a) Second harmonic pumping ($n=2$), (b) Third harmonic pumping ($n=3$), (c) Simultaneous pumping of $n=2$ and $n=3$ , and (d) Experimental observations. Note that in both cases (a) and (b), the frequency of the oscillating signal $\omega_x$ and the pump $\omega_p$ are the same, only the reference frequency $\omega_R$ (and thus $n$) are different. Small differences in frequencies for $n=2$ in (d) are due to power dependence of $\omega_x$.}
\label{figA4d}
\end{figure*}

The model predicts that for the same values of system parameters including $\omega_p$ and $\omega_x$, there could be more than one solution corresponding to different values of $n$ (thus different reference frequencies $\omega_R$). The case of multiple co-existing solutions is possible when the Nyquist criterion of stability is not fulfilled for more than one value of $n$. In this case, an effect of comb crowding (the appearance of intermediate tones) is observed. The process is shown in Fig.~\ref{figA4d}. The figure shows separate solutions for $n=2$ (a) and $n=3$ (b), and for their co-existent pumping in (c) and (d). Due to much lower magnitude instability margins of the $n=3$ solution, it is considerably less stable, thus it is observed experimentally that for the same values of the pump it may or may not appear. In the ideal situation, solutions with different $n$ are almost independent since their interaction terms are orders of magnitude smaller than the solutions themselves. The experimental observations shown in Fig. \ref{figA4d}(d) demonstrates plots for different values of the pump power, where there is a small difference between two realizations of the $n=2$ solution. This discrepancy is attributed to the power dependance of $\omega_R$ due to a self Kerr effect of the photonic mode that has not been taken into account. 

The effects observed cannot be ascribed to temperature, as temperature effects ordinarily occur over much longer time scales, especially for such a macroscopic object. In addition, no correlation between cavity temperature and sideband mid-term and long-term stability has been measured. In case of temperature-related effects, it would be natural to expect that temperature fluctuations produce long and mid-term evolution of the nonlinear system that can be detected by sideband stability measurements.

\section{Discussion and conclusion}
In this work, we related the hyperparametric phenomenon to other nonlinear observations in a model based on lumped circuits. It must be noted that the important difference with optical hyperparametric systems~\cite{savchenkov2004, matsko2005} is that optical hyperparametric oscillation (HPO) requires several near-equidistant modes with a free-spectral range (FSR) on the order of a gigahertz. Thus, they are based on nonlinear coupling between photons of different resonator modes. In our case, a comb is generated with only a single resonator mode. The present work should therefore be compared to the well-known parametric oscillators, for example, those based on high quality quartz resonators~\cite{Komine}. Such oscillators are physical implementations of the well studied Mathieu oscillators, however instead of transferring the energy from a pump which is a harmonic of the oscillating signal, the power is transferred from a pump which is detuned from the resonance frequency.

Any prior work on hyperparametric oscillations in the optical domain does not discuss the physical origin of the nonlinearity in the bulk resonator~\cite{kerr}. The origins of the Kerr nonlinearity have been studied previously in measuring contributions from the Raman process~\cite{ippen}.
It has also been demonstrated that instability phenomena in optical WG resonators are related to the effects of fast thermal relaxations in localized areas of the resonator~\cite{ilch, ilch2}. However, we rule out the possibility of temperature effects of this sort for a number of reasons. 

First, the thermal model and the fast Kerr effect~\cite{ilch2} do not predict low power drive level sensitivity, where a mode disappears at low drive level. Second, the temperature coefficient measured for the comb generating mode ($\beta = -3.9\times10^{-7}$~K$^{-1}$) is the lowest amongst all the modes characterized. 
Additionally, it is at least two orders of magnitude lower than that in case of optical WG mode resonators~\cite{ilch2}. Third, the present experiments are made at liquid helium temperatures, where temperature relaxation times are much shorter, including that of the resonator itself and the cavity. Moreover, the high quality of our temperature control is able to stabilize the resonator to within 5 mK, with no temperature increase observed. Finally, microwave WG mode resonators are significantly larger than optical microspheres, and thus the field intensity is much lower for the same amount of stored energy. Significant bulk temperature effects have been observed only down to $200$ mK where the cooling power is at minimum and no temperature control is involved. However, no parametric instability effects as described in ~\cite{ilch2} are seen.

The observed oscillatory effects cannot be explained by the interaction of a few WG modes. Predictions using method of lines of the resonator and experimental observations confirm that no other WG modes lie within the range of hyperparametric oscillations. Also, no spurious signal is observed outside this range. Additionally, the comb tunability and comb co-existence cannot be explained by the interference between modes. Similar oscillatory instabilities found in the literature on optical WG mode resonators have been ascribed to free carrier dynamics~\cite{johnson2006}. However, such polaronic effects are only physically observable in the optical range of frequencies due to the wide band-gap of rutile, which exceeds the energy of microwave frequency photons by orders of magnitude.

By applying an external magnetic field at low temperatures, the effects of paramagnetic ion frequency transitions were measured and spectroscopy was performed, allowing measurement of some ion g-factors. However, the properties of the resonances under study in this work were determined to be independent of magnetic field and were much stronger than the paramagnetic ion effects. This observation rules out the phenomena being paramagnetic.

We conclude that the nonlinear phenomena are mostly related to dielectric properties of the material. In the model proposed here, the additional dissipative branch with $\frac{1}{2}$ a degree-of-freedom is no more than a degenerate case of a coupled, highly detuned oscillator representing a material degree-of-freedom. The microwave properties of low loss materials are determined mainly from the optical domain. For example, a similar (but linear) relation between the microwave and optical domains is known to be present in sapphire resonators and other dielectrics where the permittivity and absorption coefficient for microwave modes is determined by losses in an optical phonon resonance in the infrared frequency range~\cite{LO}, where the first infrared phonon mode is even higher in frequency than in rutile. This property creates a velocity damping effect, which produces a $Q\times f = $~constant law in such systems.

\section{Acknowledgements}
This work was funded by the Australian Research Council under grants CE110001013 and FL0992016.

\end{document}